\documentclass[intlimits,twoside,a4paper]{article}

\usepackage{amsmath,amssymb}
\usepackage{graphicx}

\usepackage[T2A]{fontenc}
\usepackage[cp1251]{inputenc}

\usepackage[eqsecnum]{cmpj2}
\issue{2014}{17}{2}{23001}
\doinumber{10.5488/CMP.17.23001}

\title[A semiflexible polymer chain under geometrical restrictions]
{A semiflexible polymer chain under geometrical restrictions: Only bulk behaviour and no surface adsorption%
}

\author[P.K. Mishra]{P.K. Mishra\thanks{pkmishrabhu@gmail.com}}
\address{Department of Physics, DSB Campus, Kumaun University, Nainital-263 002 (Uttarakhand), India}

\date{Received January 20, 2014, in final form March 3, 2014}
\authorcopyright{P.K. Mishra, 2014}

\begin{document}

\maketitle

\begin{abstract}

We analyse the conformational behaviour of a linear semiflexible
homo-polymer chain confined by two geometrical constraints under a good solvent condition in two dimensions. The constraints are stair shaped impenetrable surfaces. The impenetrable surfaces are lines in a two dimensional space.
The infinitely long polymer chain is
confined in between such two ($A$ and
 $B$) surfaces. A lattice model of a fully directed self-avoiding walk is used to calculate the exact expression of the partition function, when the chain has attractive interaction
with one or both the constraints.
It has been found that under the proposed model, the chain shows only a bulk behaviour. In other words, there is no possibility of adsorption of the
chain due to restrictions imposed on the walks of the chain.
\pacs 05.70.Fh, 64.60 Ak, 05.50.+q, 68.18.Jk, 36.20.-r
\end{abstract}

\section{Introduction}

Biopolymers ({\it DNA} \textrm{and} {\it proteins}) are soft objects and, therefore, such molecules
can be easily squeezed into the spaces that are much smaller than the natural size of the molecules. For example,
{\it actin} filaments in {\it eukaryotic} cell or {\it protein} encapsulated in {\it Ecoli} \cite{1a,1b,1c} are the
examples of confined molecules that may serve as the basis for understanding molecular processes occurring in the living cells.
The conformational properties of single bio-polymers have attracted considerable attention in recent years due to
the development of single molecule based experiments \cite{2a,2b,3a,3b,3c,3d}.
The entropy of a molecule having an excluded volume interaction gets modified due to the presence of geometrical restrictions.
Therefore, geometrical constraints can modify conformational properties and the
adsorption desorption transition behaviour of the confined polymer molecules.

The behaviour of a linear and flexible polymer molecule under good solvent condition, confined to different geometries, has been studied
for the past few years \cite{4a,4b,4c,4d,5a,5b,5c,6}. Theoretical investigations of a semiflexible polymer chain under confined geometry
also find considerable attention in recent years, see \cite{7a,7b,7c,8a,8b,8c,8d,8e} and references quoted therein.
For example, Whittington and his coworkers \cite{4a,4b,4c,4d,5a,5b,5c,6} used directed self-avoiding
walk models to study the behaviour of a flexible polymer chain confined between two parallel walls on a square lattice and calculated the force
diagram for a surface interacting polymer chain. Rensburg~{et al.} \cite{6} performed numerical studies using
an isotropic self-avoiding walk model and showed that
the force diagram obtained for surface interacting polymer chains confined in between two parallel plates have a qualitatively similar
phase diagram obtained by Brak {et al.} \cite{4a,4b,4c,4d} for a directed self-avoiding walk model of the problem.

However, in the present investigation, we consider an infinitely long-linear semiflexible polymer chain confined in between one dimensional
two stair shaped impenetrable surfaces (geometrical constraints) under good solvent conditions and we discuss
the conformational behaviour of the chain.
Such an investigation may be useful to understand the behaviour of a macromolecule near a membrane as well as the behaviour of
{\it DNA} in micro-arrays and electrophoresis.

\looseness=-1
To analyze the conformational behaviour of such semiflexible chains we have chosen a
fully directed self-avoiding walk model introduced by Privmann and coworkers \cite{9a,9b} and have used a generating function technique
to solve the model analytically for different values of the spacing between the constraints.
The result so obtained is used to discuss the possibility of an adsorption phase transition behaviour of the polymer chain on the stair
shaped geometrical constraints.
Since the constraint is an attractive surface, it contributes an energy $\epsilon_\mathrm{s}$
($<0$) for each step of the fully directed self-avoiding walk touching the constraint. This leads
to an increased probability defined by a Boltzmann weight $\omega=\exp(-\epsilon_\mathrm{s}/k_\mathrm{B}T)$ of stepping on the constraint
($\epsilon_\mathrm{s} < 0$ or $\omega > 1$, $T$ is temperature and $k_\mathrm{B}$ is the Boltzmann constant).
The polymer chain gets adsorbed on the constraint at an appropriate  value of $\omega$ or $\epsilon_\mathrm{s}$.
Therefore, the transition between an adsorbed to a desorbed phase is marked by a critical value of adsorption
energy $\epsilon_\mathrm{s}$ or $\omega_\mathrm{c}$.

In this paper, we analytically solve the fully directed self-avoiding walk model to calculate the exact expression of the partition function of the chain
when the chain has an attractive interaction either with one or both of the geometrical constraints.
The results so obtained are compared with the
case when the adsorption of a semiflexible
polymer chain occurs on a flat surface \cite{10a,10b,10c,10d,10e,10f,11}.

The paper is organized as follows: In section~2, a square-lattice model of
fully directed self-avoiding walk is described for an infinitely long and linear semiflexible homo-polymer chain confined in between the constraints
for a given value of spacing between the constraints.
In subsection~2.1, we discuss the possibility of an adsorption transition
of the polymer chain when constraint $A$ has an attractive interaction with the semiflexible polymer chain.
Subsection~2.2 is devoted to a discussion of the adsorption of a semiflexible polymer chain on the constraint $B$.
While in subsection~2.3 the expression of the partition function of the polymer chain is obtained for the case when the
chain has an attractive interaction with both the constraints.
Finally, in section~3 we summarize and discuss the results obtained.

\section{Model and method}

\begin{figure}
\centerline{
\includegraphics[width=0.9\textwidth]{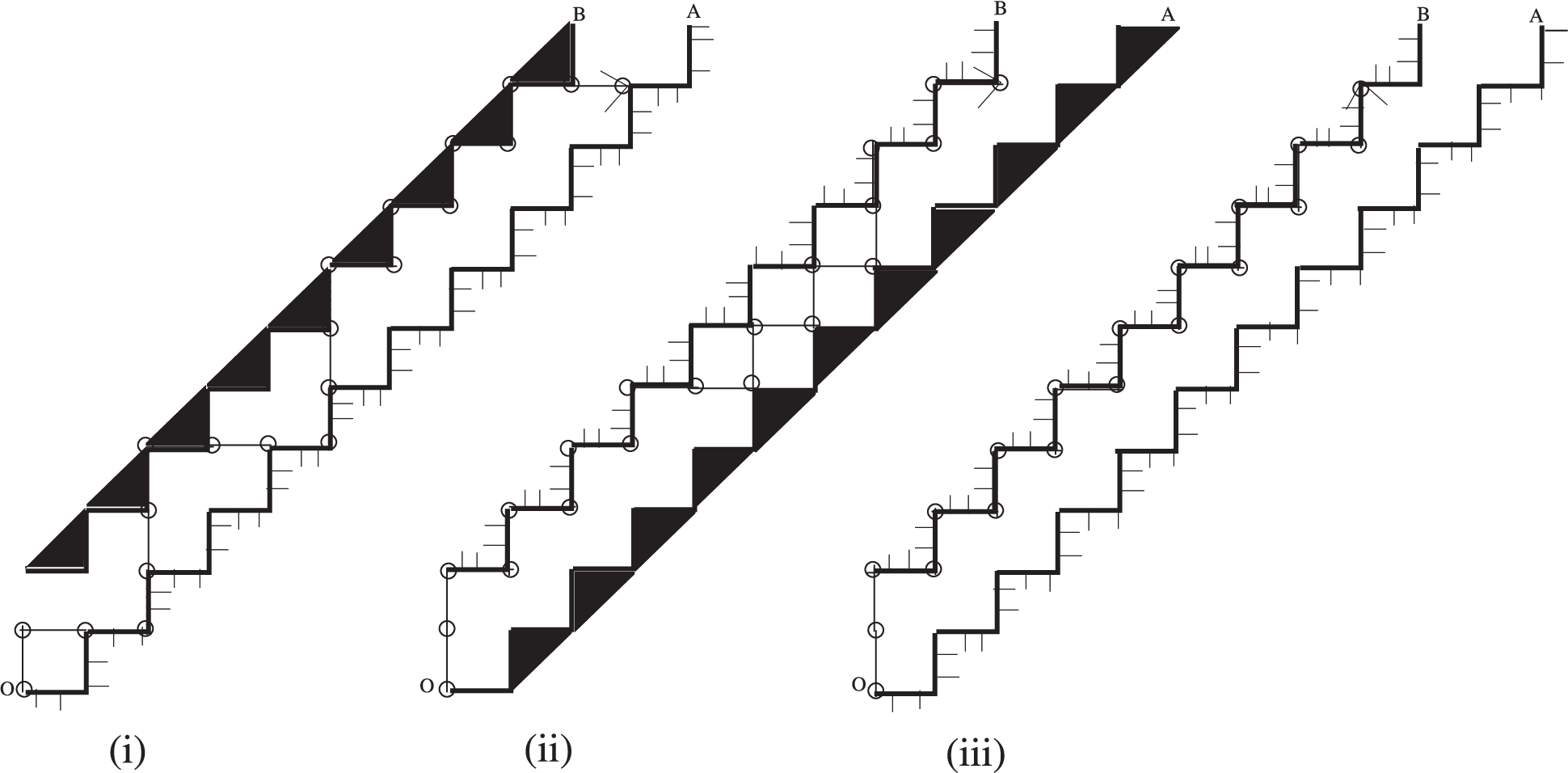}
}
\caption{This figure shows a walk of an infinitely long linear semiflexible polymer chain confined in between two constraints
(impenetrable stair-shaped surface). All walks of the chain start from a point $O$ on the constraint.
We show three different cases viz. (i), (ii) and (iii) having separation ($n$) between the
constraints along the axis three monomers (steps). The separation between the constraints are defined
on the basis of how many steps a walker can successively move at maximum along any of the $+x$ or $+y$ directions.
In the case 1~(i), the constraint $A$ has an attractive interaction with the monomers of the chain, in 1~(ii) only constraint $B$ has an attractive
interaction with the monomers of the chain while in 1~(iii) both constraints are shown to have an attractive interaction with the monomers
of the polymer chain.}
\label{fig:1}       % Give a unique label
\end{figure}

A model of fully directed self-avoiding walks \cite{9a,9b} on a square lattice is used to
investigate the possibility of an adsorption transition of an infinitely long linear semiflexible homopolymer chain
on geometrical constraints, when the chain is confined in between two
impenetrable stair shaped surfaces under a good solvent condition (as shown schematically in figure~\ref{fig:1}).  The directed walk model is
restrictive in the sense that the angle of bending has a unique value, that is $90^{\circ}$ for a square lattice
and the directivity of the walk amounts to a certain degree of stiffness in the
walks of the chain because different
directions of the space are not treated equally.
Since the directed self-avoiding walk model can be solved
analytically, it gives exact values of the
partition function of the polymer chain.
We consider a fully directed self-avoiding walk (FDSAW)
model. Therefore, the walker is allowed to take steps along $+x$, and $+y$ directions on a square lattice in between the constraints.
%\vspace{1cm}

The walks of the chain start from a point $O$ located on the impenetrable surface-$A$, and the walker moves in the space in between the
two surfaces [as we have shown schematically in figure~\ref{fig:1}, a walk of the polymer chains confined in between two surfaces for a value
of separation $n(=3$) between them].

The stiffness of the chain is accounted for by associating a Boltzmann weight with the bending energy for each turn in the walk of the polymer chain.
The stiffness weight is $k${\Large}$=\exp(-\beta\epsilon_\mathrm{b})$; where $\beta={1}/{k_\mathrm{B}T}$ is the
inverse of the temperature, $\epsilon_\mathrm{b}(>0)$ is the energy associated with
each bend in the walk of the chain, $k_\textrm{B}$ is the Boltzmann constant and $T$ is temperature{\Large}.
For $k=1$ or $\epsilon_\mathrm{b}=0$, the chain is said to be flexible and for
$0<k<1$ or $0<\epsilon_\mathrm{b} <\infty$ the polymer chain is said to be
semiflexible. However, when $\epsilon_\mathrm{b}\to\infty$ or $k \to 0$,
the chain has the shape of a rigid rod.

The partition function of a surface interacting semiflexible polymer chain can be written as follows:
\begin{equation}
\label{eq:2.1}
Z(\omega,k)={\sum}^{N=\infty}_{N=0}
\sum_{ \text{all walks of $N$ steps} } {g^N\omega}^{N_\mathrm{s}}k^{N_\mathrm{b}}\, ,
\end{equation}
where $N_b$ is the total number of bends in a walk of $N$ steps (monomers), $N_\mathrm{s}$ is number of monomers in a $N$ step walk ($N_\mathrm{b} \leqslant N-1$ and $N_\mathrm{s} \leqslant N$),
lying on the surface,
$g$ is the step fugacity of each monomer of the chain, and $\omega$ is the Boltzmann weight of the monomer-surface attraction energy.

\subsection{A semiflexible polymer chain interacting with constraint $A$}
\label{sec:1}

\begin{figure}
\centerline{
\includegraphics[width=0.8\textwidth]{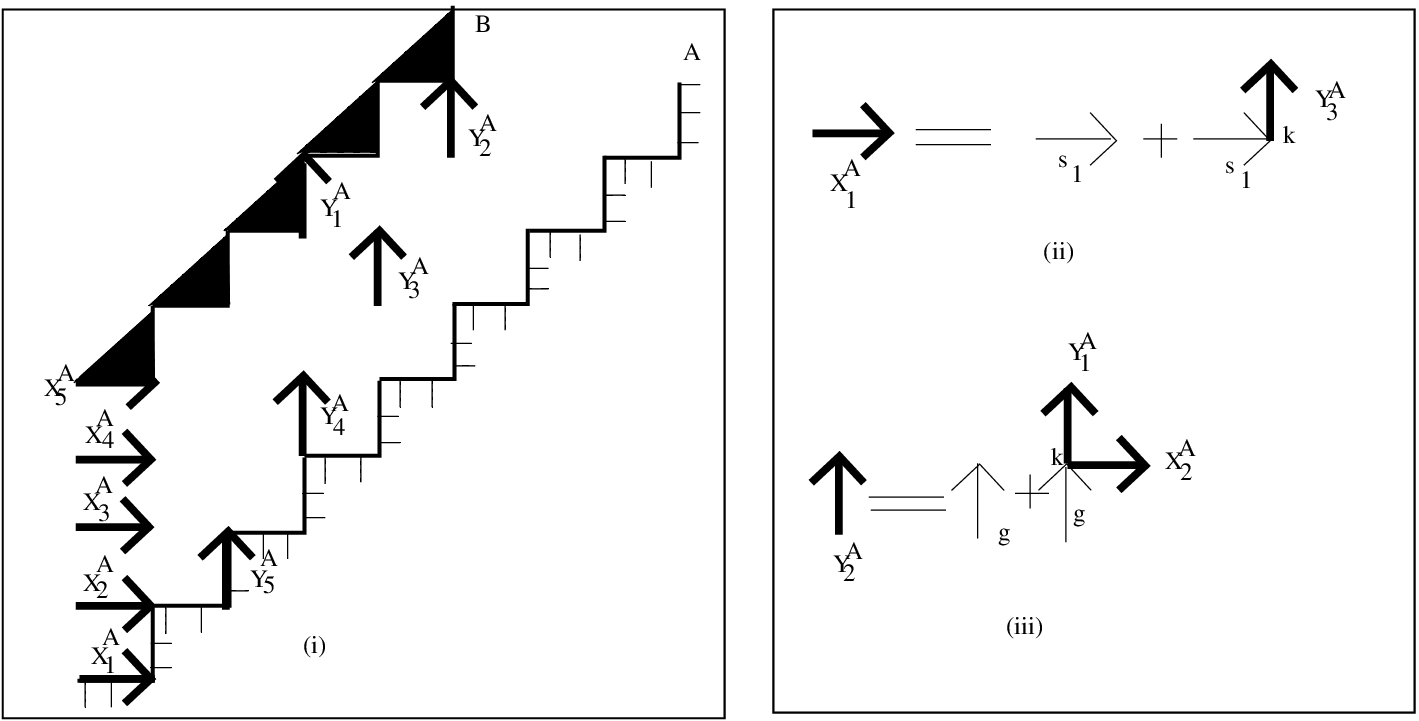}
}
\caption{The components of the partition function is shown graphically in this figure. Term $X_m^A$ ($3\leqslant m\leqslant n$) indicates the sum of Boltzmann weight
of all the walks having the first step along $+x$ direction and suffix $n$ indicates maximum number of steps that a walker can successively take
along $+x$ direction. Similarly, we have defined $Y_m^A$, where the first step of the walker is along $+y$ direction. In this figure, (ii) and (iii)
graphically represents the recursion relation for equations (\ref{eq:2.2}) and (\ref{eq:2.6}), respectively.}
\label{fig:2}
\end{figure}

The partition function of an infinitely long linear semiflexible polymer chain confined in between the constraints
(as shown schematically in figure~\ref{fig:1}~(i) and having an attractive interaction with the constraint $A$ can be calculated using the
generating function technique. The components (as shown in figure~\ref{fig:2}) of the partition function
$Z_3^A(k,\omega_1)$ (here we have used the suffix three because in figure~\ref{fig:1}~(i) case, the maximum step that a walker can move
successively in one particular direction is three and $\omega_1$ is the Boltzmann weight of the attraction energy between monomers, and thus the
constraint $A$) of the chain can be written as follows:
\begin{equation}
\label{eq:2.2}
X_1^A=s_1+ks_1Y_3^A,
\end{equation}
where $s_1=\omega_1g$.
\begin{align}
\label{eq:2.3}
X_2^A &=g+g\left(X_1^A+kY_2^A\right),\\
%\end{equation}
%
%\begin{equation}
\label{eq:2.4}
X_3^A&=g+g\left(X_2^A+kY_1^A\right),\\
%\end{equation}
%
%\begin{equation}
\label{eq:2.5}
Y_1^A&=g+kgX_3^A,\\
%\end{equation}
%
%\begin{equation}
\label{eq:2.6}
Y_2^A&=g+g\left(kX_2^A+Y_1^A\right),
\end{align}
and
\begin{equation}
\label{eq:2.7}
Y_3^A=s_1+s_1\left(kX_1^A+Y_2^A\right).
\end{equation}

On solving equations (\ref{eq:2.2})--(\ref{eq:2.7}), we find the expression for $X_1^A(k,\omega_1)$ and $Y_2^A(k,\omega_1)$. In obtaining the expression for $X_1^A(k,\omega_1)$
and $Y_2^A(k,\omega_1)$,
we have solved a matrix of $2n\times2n$ ($n=3$), for the present case {i.e.}, figure~\ref{fig:1}~(i).
Thus, we have an exact expression of the partition function for an infinitely long linear semiflexible polymer chain confined between the constraints
and having an attractive interaction with the constraint $A$ [as shown in ~\ref{fig:1}~(i)]. This is written as follows:
\begin{equation}
\label{eq:2.8}
Z_3^A(k,\omega_1)=X_1^A(k,\omega_1)+Y_2^A(k,\omega_1)
=-\frac{u_1+u_2+u_3+2k^4s_1^2g^4-2k^5s_1^2g^4}{-1+k^6s_1^2g^4+u_4}\,,
\end{equation}
where
\begin{align*}
u_1&=s_1-ks_1^2-g-ks_1^2g+k^2s_1^2g-g^2-kg^2-ks_1g^2+2k^2s_1g^2-ks_1^2g^2,\\
u_2&=-k^2s_1^2g^2+3k^3s_1^2g^2-kg^3+k^2g^3-k^2s_1^2g^3+2k^3s_1^2g^3-k^4s^2g^3,\\
u_3&=-kg^4+k^3g^4-ks_1g^4+k^2s_1g^4+k^3s_1g^4-k^4s_1g^4-2k^2s_1^2g^4+2k^3s_1^2g^4,
\end{align*}
and
\[
u_4=-k^4\left[g^4+2s_1^2\left(g^2+g^4\right)\right]
+k^2\left[g^2\left(2+g^2\right)+s_1^2\left(1+g^2+g^4\right)\right].
\]

From the singularity of the partition function, $Z_3^A(k,\omega_1)$, we obtain the critical value of the Boltzmann's weight for the
monomer-constraint $A$ attraction energy,
\[
\omega_{\mathrm{c}1}=\frac{\sqrt{1-2k^2g^2-k^2g^4+k^4g^4}}{\sqrt{k^2g^2+k^2g^4-2k^4g^4+k^2g^6-2k^4g^6+k^6g^6}}\,.
\]

This is required for the
adsorption of an infinitely long linear semiflexible polymer chain on the constraint $A$.
We obtain the value of $\omega_{\mathrm{c}1}=1$, when we substitute the value of $g_\mathrm{c}$ in the expression of $\omega_{\mathrm{c}1}$ corresponding to all possible values
of $k[=\exp(-\beta\epsilon_\mathrm{b})]$ or the bending energy $\epsilon_\mathrm{b}$ for which an infinitely long linear semiflexible polymer chain can be
polymerized in between the constraints. It shows the existence of only one singularity $g_\mathrm{c}$ of the partition function equation~(\ref{eq:2.8})
and it corresponds to the bulk behaviour of the chain.
There is no possibility of an adsorption transition of the chain on constraint $A$.

\subsection{A semiflexible polymer chain interacting with constraint $B$}
\label{sec:2}
The partition function of an infinitely long linear semiflexible polymer chain confined in between the constraints
[as shown schematically in figure~\ref{fig:1}~(ii)] and having an attractive interaction with the constraint $B$ is calculated following the
method discussed in the above subsection. The components of the partition function
$Z_3^B(k,\omega_2)$ (where $\omega_2$ is Boltzmann weight of attraction energy between the monomers of the chain and the
constraint $B$) of the chain can be written as follows:
\begin{align}
\label{eq:2.9}
X_1^B&=g+kgY_3^B,\\
%\end{equation}
%\begin{equation}
\label{eq:2.10}
X_2^B&=g+g\left(X_1^B+kY_2^B\right),\\
%\end{equation}
%\begin{equation}
\label{eq:2.11}
X_3^B&=s_2+s_2\left(X_2^B+kY_1^B\right),
\end{align}
where $s_2=\omega_2g$.
\begin{align}
\label{eq:2.12}
Y_1^B&=s_2+ks_2X_3^B,\\
%\end{equation}
%\begin{equation}
\label{eq:2.13}
Y_2^B&=g+g\left(kX_2^B+Y_1^B\right),
\end{align}
and
\begin{equation}
\label{eq:2.14}
Y_3^B=g+g\left(kX_1^B+Y_2^B\right).
\end{equation}

On solving equations (\ref{eq:2.9})--{\ref{eq:2.14}), we find an expression for $X_1^B(k,\omega_2)$ and $Y_2^B(k,\omega_2)$. In obtaining the expression for $X_1^B(k,\omega_2)$
and $Y_2^B(k,\omega_2)$,
we have to solve a matrix of $2n \times 2n$ [$n=3$, for figure~\ref{fig:1}~(ii) case].
Thus, we obtain an exact expression of the partition function for an infinitely long linear semiflexible polymer chain confined between the constraints
and having an attractive interaction with the constraint $B$ [as shown in figure~\ref{fig:1}~(ii)] which is as follows:
\begin{equation}
\label{eq:2.15}
Z_3^B(k,\omega_2)=X_1^B(k,\omega_2)+Y_2^B(k,\omega_2)
=-\frac{-g(s_2(1+kg^2-k^2g^2)+2u_5+ks_2^2u_6)}{-1+k^6s_2^2g^4+u_7}\,,
\end{equation}
where
\begin{align*}
u_5&=1+k^2(-1+g)g^2-k^3g^3+kg(1+g),\\
u_6&=1+g+g^2-2k^3(-1+g)g^2+2k^4g^3-k^2g\left(2+3g+3g^2\right)+2k\left(-1+g^3\right),\\
u_7&=-k^4\left[g^4+2s_2^2\left(g^2+g^4\right)\right]+k^2\left[g^2\left(2+g^2\right)
+s_2^2\left(1+g^2+g^4\right)\right].
\end{align*}

From the singularity of the partition function, $Z_3^B(k,\omega_2)$, we obtain a critical value for the monomer-constraint $B$ attraction energy,
\[
\omega_{\mathrm{c}2}=\frac{\sqrt{1-2k^2g^2-k^2g^4+k^4g^4}}{\sqrt{k^2g^2+k^2g^4-2k^4g^4+k^2g^6-2k^4g^6+k^6g^6}}=\omega_{\mathrm{c}1},
\]
required for
adsorption of an infinitely long linear semiflexible polymer chain on the constraint $B$.
In this case too, we find $\omega_{\mathrm{c}2}=1$, for all possible values of the bending energy or stiffness of the semiflexible polymer chain and
further there is no possibility for the existence of a new singularity of the partition function {i.e.} equation (\ref{eq:2.15}).
Therefore, the adsorption of the chain on constraint $B$
is impossible.

\subsection{A semiflexible polymer chain interacting
with both the constraints $A \ \textrm{and}  \  B$}
\label{sec:3}

The partition function of an infinitely long linear semiflexible polymer chain confined in between the constraints
[as shown schematically in figure~\ref{fig:1}~(iii)] and having an attractive interaction with both the constraints ($A \ \textrm{and}  \ B$) is calculated following the
method discussed in the above subsections. The components of the partition function
$Z_3^\mathrm{C}(k,\omega_3,\omega_4)$ of the chain can be written as follows:
\begin{equation}
\label{eq:2.16}
X_1^\mathrm{C}=s_3+ks_3Y_3^\mathrm{C},
\end{equation}
where $s_3=\omega_3g$.
\begin{align}
\label{eq:2.17}
X_2^\mathrm{C}&=g+g\left(X_1^\mathrm{C}+kY_2^\mathrm{C}\right),\\
%\end{equation}
%\begin{equation}
\label{eq:2.18}
X_3^\mathrm{C}&=s_4+s_4\left(X_2^\mathrm{C}+kY_1^\mathrm{C}\right),
\end{align}
here, $s_4=\omega_4g$.
\begin{align}
\label{eq:2.19}
Y_1^\mathrm{C}&=s_4+ks_4X_3^\mathrm{C},\\
%\end{equation}
%\begin{equation}
\label{eq:2.20}
Y_2^\mathrm{C}&=g+g\left(kX_2^\mathrm{C}+Y_1^\mathrm{C}\right),
\end{align}
and
\begin{equation}
\label{eq:2.21}
Y_3^\mathrm{C}=s_3+s_3(kX_1^\mathrm{C}+Y_2^\mathrm{C}) .
\end{equation}

On solving equations~(\ref{eq:2.16})--(\ref{eq:2.21}), we get the expression for $X_1^\mathrm{C}(k,\omega_3,\omega_4)$ and $Y_2^\mathrm{C}(k,\omega_3,\omega_4)$. In obtaining the expression
for $X_1^\mathrm{C}(k,\omega_3,\omega_4)$ and $Y_2^\mathrm{C}(k,\omega_3,\omega_4)$, we have solved a matrix of $2n \times 2n$ [$n=3$, for figure~\ref{fig:1}~(iii) case].
Thus, we have an exact expression for the partition function of an infinitely long linear semiflexible polymer chain confined between the constraints
and having an attractive interaction with the constraints [as shown in figure~\ref{fig:1}~(iii)]. This is written as follows:
\begin{equation}
\label{eq:2.22}
Z_3^\mathrm{C}(k,\omega_3,\omega_4)=X_1^\mathrm{C}(k,\omega_3,\omega_4)
+Y_2^\mathrm{C}(k,\omega_3,\omega_4)
=-\frac{-(gu_8+s_3u_9)+u_{10}+u_{11}}{-1+k^6s_3^2s_4^2g^2+u_{12}+u_{13}}\,,
\end{equation}
where
\begin{align*}
u_8&=1+s_4+kg-(-1+k)ks_4^2(1+g+kg),\\
u_9&=1-k^2s_4^2g^2+k^4s_4^2g^2+k\left(1+s_4^2\right)g^2-k^2\left[g^2s_4^2\left(1+g^2\right)\right],\\
u_{10}&=ks_3^2\left[1+g+s_4gk^2s_4^2\left(1-2g\right)g+2k^4s_4^2g^2+kg\left(-1-s_4+2g\right)\right],\\
u_{11}&=ks_3^2\left\{kgs_4^2(1+2g)-k^2\left[2g^2+s_4^2\left(1+2g+2g^2\right)\right]\right\},\\
u_{12}&=k^2\left\{g^2+s_4^2\left(1+g^2\right)+s_3^2\left[1+\left(1+s_4^2\right)g^2\right]\right\},
\end{align*}
and
\[
u_{13}=-k^4\left\{s_4^2g^2+s_3^2\left[g^2+s_4^2\left(1+2g^2\right)\right]\right\}.
\]

From the singularity of the partition function, $Z_3^\mathrm{C}(k,\omega_3,\omega_4)$, we obtain a critical value of the monomer-constraint attraction energy,
\begin{equation}
\label{eq:2.23}
\omega_{\mathrm{c}3}=\frac{\sqrt{1-k^2g^2-k^2g^2\omega_4^2-k^2g^4\omega_4^2+k^4g^4\omega_4^2}}{\sqrt{k^2g^2+k^2g^4-k^4g^4-k^4g^6\omega_4^2
-2k^4g^6\omega_4^2+k^2g^6\omega_4^2+k^6g^6\omega_4^2}}\,,
\end{equation}
required for the
adsorption of an infinitely long linear semiflexible polymer chain on the constraints $A$, when both the constraints have an attractive
interaction with the chain.

On substitution of the value of $\omega_4$ in equation (\ref{eq:2.23}) to get the value of $\omega_{\mathrm{c}3}=1$,
\[
\omega_{\mathrm{c}4}=\frac{\sqrt{1-2k^2g^2-k^2g^4+k^4g^4}}
{\sqrt{k^2g^2+k^2g^4-2k^4g^4+k^2g^6-2k^4g^6+k^6g^6}}=\omega_{\mathrm{c}2}\,.
\]

The method discussed above can be used for different values of $n$. The size of the matrix needed to solve for the partition function of the chain
confined in between the constraints is $2n \times 2n$. We have calculated the exact expressions of the partition function for $n$ $(3\leqslant n \leqslant 19)$.

We have found that the adsorption transition point of an infinitely long linear semiflexible polymer chain on the constraint $A$, $B$ and simultaneously
on both the constraints $A\ \textrm{and}  \ B$ has the value of unity. The equation (\ref{eq:2.23}) has only a singularity that corresponds to the polymerization of an infinitely long
linear homopolymer chain in between the constraints. Therefore, there is no possibility of the adsorption-desorption phase transition in
the proposed model.
This fact is true for the chosen values of $k$ or the bending energy (as checked for $3\leqslant n\leqslant\ 19$) for which
an infinitely long polymer chain can be polymerized in between the constraints.

\subsection{General expressions of the recursion relations}
In this subsection, we should like to express the recursion relations with the least possible number of equations. This method is useful in solving a matrix
of $n\times n$ rather than $2n\times 2n$ as discussed in the subsections 2.1--2.3. For instance, equations (\ref{eq:2.16})--(\ref{eq:2.21}) can be written as follows:
\begin{eqnarray}
\label{eq:2.24}
W^n_1&=&s_3+ks_3^2+kgs_3^3+\dots+kg^{n-2}s_3^2s_4+k^2s_3^2W_1^n \nonumber \\
&&+k^2gs_3^2W_2^n+k^2g^2s_3^2W_3^n+\dots+k^2g^{n-2}s_3^2s_4W_n^n,\\
%\end{eqnarray}
%\begin{eqnarray}
\label{eq:2.25}
W_m^n&=&g+kg^2+\dots+kg^{n+1-m}s_4+gW_{m-1}^n+k^2g^2W_m^n \nonumber \\
&&+k^2g^3W_{m+1}^n+ \dots+k^2g^{n+1-m}W_{n-1}^n+k^2g^{n+1-m}s_4W_n^n,
\end{eqnarray}
where $(1<m<n)$ and $W_m^n=0$, when $m<1$.
\begin{equation}
\label{eq:2.26}
W_n^n=s_4+ks_4^2+s_4W_{n-1}^n+k^2s_4W_n^n.
\end{equation}
The equations (\ref{eq:2.24})--(\ref{eq:2.26}) can be used to express recursion relations, $X_n^\mathrm{C}$ (for all values of the chosen $n$), and mutual exchange of $s_3$ with $s_4$
will result in the recursion relations $Y_n^\mathrm{C}$ for the chosen values of $n$. The partition function of the chain can now be written as follows:

\begin{equation}
\label{eq:2.27}
Z_n^\mathrm{C}(k,\omega_3,\omega_4)=W^n_1+W_{n-1}^n,
\end{equation}
where $W^n_1$ is the sum of the Boltzmann weights of all walks starting from a point $O$ lying on the constrant $A$ and
having the first step along $+x$ direction,
while $W_{n-1}^n$ is the sum of the Boltzmann weights of all the walks starting from point $O$ and with the first step along $+y$ direction.

However, substituting $s_4=g$ and $s_3=s_1$, we have recursion relations and a partition function for the case 1~(i), as shown in figure~\ref{fig:1} and when we
substitute $s_3=g$ and $s_4=s_2$, recursion relations and partition function for the case 1~(ii) of figure~\ref{fig:1} were found by us.

If the constraints are assumed to be neutral, the recursion relations can be written for any given value of $n$ as follows:
\begin{eqnarray}
\label{eq:2.28}
W_m^n&=&g+kg^2+kg^3+\dots+kg^{n+2-m}+gW_{m-1}\nonumber\\
&&+k^2g^2W_m+k^2g^3W_{m+1}+\dots+k^2g^{n+2-m}W_n^n,
\end{eqnarray}
where $1\leqslant m\leqslant n$ and $W_0^n=0$.

\section{Summary and conclusions}
We have considered an infinitely long linear semiflexible homopolymer chain confined in between two impenetrable stair shaped
surfaces (constraint) in two dimensions under good solvent condition. We have used a fully directed self-avoiding walk
model to study the adsorption phase transition behaviour of the polymer chain on any of the two constraints ($A$  \textrm{and}  $B$)
and simultaneous adsorption of the polymer chain on both the constraints ($A$  \textrm{and}  $B$). The generating function technique is used
to solve the model analytically and an exact expression of the partition function of the surface interacting semiflexible polymer chain
is obtained for different values of spacing $(3\leqslant n \leqslant 19)$ between the constraints.

We find in the case 1~(i), 1~(ii) and 1~(iii) that the bulk behaviour of the polymer
chain occurs on the constraints for the values $\omega_{\mathrm{c}1}=\omega_{\mathrm{c}2}=\omega_{\mathrm{c}3}=1$ for all possible values of $k$ or the bending energy of the chain for
which an infinitely long linear semiflexible polymer chain can be polymerized in between the constraints. The critical value of $\omega$
is unity for all cases considered and for different values of spacing between the constraints ($3\leqslant n\leqslant19$). This result is
obvious because the walks of the chain are directed along the constraint(s), therefore, the partition function of the chain is dominated by the walks
lying on the constraints, and the bulk behaviour is observed on the constraints. We have
shown the results for a few values of $n=3,\,7,\,11,\,16$ in the table~\ref{tab1} for the case 1~(i), when the chain  interacts with the constraint $A$.
The chain is grafted to
the constraint $A$ for the case 1~(i), 1~(ii) and 1~(iii), as shown in figure~\ref{fig:1}. An infinitely long linear chain is polymerized in between the two
constraints $A$ and $B$, when $g=g_\mathrm{c}$.

However, in the case of adsorption of an infinitely long linear semiflexible polymer chain on a flat surface, the adsorption transition point
is found to depend on the bending energy or stiffness of the chain. In this case, the partition function of the surface interacting chain has two
singularities. One singularity corresponds to the bulk behaviour {i.e.,} polymerization of an infinitely long linear chain and the other
singularity corresponds to adsorption transition of the chain on the surface \cite{10a,10b,10c,10d,10e,10f,11}.

We have also expressed general expressions of the recursion relations, when the chain has an attractive interaction with any or both the
constraints and when the constraints are assumed to be neutral. It has been found that polymerization of an infinitely long flexible polymer
chain is not possible for separations ($n$) between constraints 3, 6 and 8. In the case of $n=3$, the imaginary part of the critical value of step fugacity
is negligible. However, for other values of separation between the constraints, {i.e.,}
$n=6$ and 8 the imaginary part in the critical value of step fugacity is reasonable and cannot be ignored.
We plan to discuss these issues in a another paper to be submitted elsewhere in due time.

\begin{table}[htb]
\caption{This table shows the values of $g_\mathrm{c}$ and $s_\mathrm{c}(=\omega_{\mathrm{c}1}g_\mathrm{c})$ for different values of separation ($n$) between the constrains, for the case 1~(i), as shown in figure~\ref{fig:1}. The value of $s_\mathrm{c}=g_\mathrm{c}$ indicates that $\omega_{\mathrm{c}1}=1$.}
\label{tab1}
\vspace{2ex}
\begin{center}
\begin{tabular}{|c|c|c|c|c|c|c|c|c|}
\hline\hline
&\multicolumn{2}{|c|}{$n=3$}&\multicolumn{2}{|c|}{$n=7$}&\multicolumn{2}{|c|}{$n=11$}
&\multicolumn{2}{|c|}{$n=16$} \\
\hline
$k$ &  $g_\mathrm{c}$ & $s_\mathrm{c}$ & $g_\mathrm{c}$& $s_\mathrm{c}$ & $g_\mathrm{c}$ & $s_\mathrm{c}$ & $g_\mathrm{c}$ & $s_\mathrm{c}$  \\
\hline\hline
$0.1$ &  $-$& $-$ &  $-$ & $-$ & $-$ & $-$ &  $0.99303$ & $0.99303$  \\
$0.2$  & $-$ & $-$  &  $-$ & $-$  &   $0.92215$ & $0.92215$  &   $0.88185$ & $0.88185$  \\
$0.3$ &  $-$ & $-$  &   $0.89545$ & $0.89545$  &   $0.83077$ & $0.83077$  &   $0.80227$ & $0.80227$  \\
$0.4$ &   $-$ & $-$  &  $0.80962$ & $0.80962$  &  $0.76023$ & $0.76023$  &   $0.73867$ & $0.73867$  \\
$0.5$ &  $0.93879$ & $0.93879$ &   $0.74186$ & $0.74186$  &   $0.70259$ & $0.70259$ &  $0.68558$ & $0.68558$  \\
$0.6$ &   $0.84709$ & $0.84709$  &  $0.68613$ & $0.68613$  &  $0.65400$ & $0.65400$ &  $0.64017$ & $0.64017$ \\
$0.7$ &  $0.77358$ & $0.77358$  &  $0.63907$ & $0.63907$  &   $0.61222$ & $0.61222$ &   $0.60072$ & $0.60072$  \\
$0.8$ &  $0.71293$ & $0.71293$  &  $0.59857$ & $0.59857$  &  $0.57576$ & $0.57576$ &   $0.56603$ & $0.56603$ \\
$0.9$ &  $0.66182$ & $0.66182$  &   $0.56325$ & $0.56325$  &   $0.54359$ & $0.54359$ &   $0.53524$ & $0.53524$  \\
$1.0$ &  $0.61803$ & $0.61803$  &  $0.53208$ & $0.53208$  &   $0.51496$ & $0.51496$ &  $0.50771$ & $0.50771$ \\
\hline\hline
\end{tabular}
\end{center}
\end{table}

%\makeukrtitle
% For tables use
%\begin{acknowledgements}
\section*{Acknowledgements}

The financial support received from Department of Science and Technology, New Delhi (SR/FTP/PS-122/2010)
thankfully acknowledged. The author also would like to thank
Professor D. Dhar, TIFR, Mumbai (India) and the anonymous referee for useful corrections in the earlier version of the manuscript.

%% BibTeX users please use one of
%\bibliographystyle{aps-nameyear}      % American Physical Society (APS) style, author-year citations
%\bibliography{example}                % name your BibTeX data base
%\nocite{*}

\ukrainianpart

\title{Напівгнучкий полімерний ланцюг під дією геометричних обмежень: лише  об'ємна поведінка і жодної поверхневої адсорбції}

\author{П.К. Мішра}
\address{Фізичний факультет, Кумаонський університет, м. Найнітал (Уттаракханд), Індія}

\makeukrtitle

\begin{abstract}
\tolerance=3000%
Ми аналізуємо конформаційну поведінку лінійного напівгнучкого гомополімерного
ланцюга під дією двох геометричних обмежень
в умовах доброго розчинника у двовимірному просторі. Обмеження представляють собою
непроникні поверхні східчастої форми. Непроникні поверхні
є лініями у двовимірному просторі.  Нескінченно довгий полімерний ланцюг є обмежений
двома  поверхнями ($A \ \textrm{and}  \ B$).  Для розрахунку точного виразу статистичної суми
використовується ґраткова модель повністю напрямлених блукань без самоперетинів,
якщо ланцюг має притягальну взаємодію з одною чи двома поверхнями. В рамках  запропонованої моделі
отримано лише об'ємну поведінку  ланцюга. Іншими словами, жодної можливості для адсорбції ланцюга
під дією обмежень на блукання, не спостерігається.

\keywords полімерна адсорбція, об'ємна поведінка, геометричні обмеження, точні результати

\end{abstract}

\end{document}